\documentclass[twocolumn,showpacs,amsmath,amssymb]{revtex4}
\usepackage{dcolumn}

\usepackage{graphicx}

\begin{document}

\title{Reply to the comment on ``Incomplete equilibrium in long-range interacting
    systems'' by Tsallis {\it et al.}}

\author{Fulvio Baldovin and Enzo Orlandini}
\email{baldovin@pd.infn.it, orlandini@pd.infn.it}
\affiliation{
Dipartimento di Fisica and
Sezione INFN, Universit\`a di Padova,\\
\it Via Marzolo 8, I-35131 Padova, Italy
}

\date{April 3, 2007}

\begin{abstract}
After the rejection of their comment [arXiv:cond-mat/0609399v1] to our
Phys. Rev. Lett. {\bf 97}, 100601 (2006), the Authors informed us that
an extended version of their comment is going to be published in a
different journal under the direct editorial responsibility of one of
them. We then decided to make publicly available our formal reply,
originally prepared for publication in Phys. Rev. Lett.
\end{abstract}


\maketitle

In their comment \cite{tsallis}, Tsallis {\it et al.} present three
points which they claim to confute the conclusions of our work in
Ref. \cite{hmf_qss}. Here we show that this is not the case.  

An important issue about the 
quasi-stationary states (QSS's) displayed by isolated long-range
interacting systems before relaxing to equilibrium 
is whether the QSS's survive to the
perturbations introduced by a thermal reservoir (canonical QSS's). 
In order to answer this question, we proposed \cite{hmf_can} 
an Hamiltonian setup in which 
the microcanonical conditions are recovered 
when the coupling constant between system and thermal bath
vanishes. 
Indeed, we showed \cite{hmf_can} that 
canonical QSS's exist, with a lifetime which decreases as the
coupling strength increases. 
In \cite{hmf_qss} we discuss the statistics of the system energy
fluctuations and, contrarily to what is claimed in \cite{tsallis},
we do not  
``\ldots extrapolate the conclusions for the canonical QSS's \ldots to
the microcanonical ones'', where by definition the energy fluctuations
are zero.  
Hence, the first criticism raised in \cite{tsallis} 
offers to the unaware Reader 
a false representation of our results and motivations.

\begin{figure}
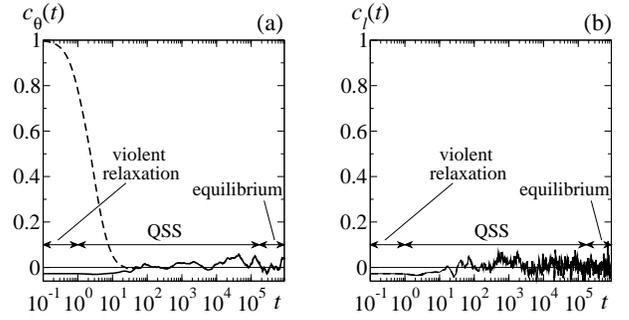

\vspace*{12pt}
\includegraphics[width=0.49\columnwidth]{corr_theta.eps}
\includegraphics[width=0.49\columnwidth]{corr_vel.eps}
\caption{
Particle-particle correlation during a microcanonical simulation of the
Hamiltonian system in \cite{hmf_qss} with $M=5000$ particles. 
$C_x$ ($x\equiv\theta,l$) is defined as 
$C_x\equiv(\langle x_i x_{i+I}\rangle-\langle x_i\rangle \langle x_{i+I}\rangle)/
\sqrt{(\langle x_i^2\rangle-\langle x_i\rangle^2)
(\langle x_{i+I}^2\rangle-\langle x_{i+I}\rangle^2})
$ 
(full line) and 
$C_x\equiv\langle x_i x_{i+I}\rangle/
(\sqrt{\langle x_i^2\rangle
\langle x_{i+I}^2\rangle})
$ (dashed line), with $\langle\cdot\rangle\equiv(\sum_{i=1}^I\cdot)/I$ and
$I<M/2$.  
Since at $t=0$ all angles $\theta_i$'s coincide (initial
magnetization is $1$), in (a) the dashed line starts from
$c_\theta=1$.  
In (b) the angular momenta $l_i$'s are initially uniformly distributed and
the dashed and the full lines coincide for all times. 
In all simulations we observed that $C_x$ during the QSS is not
appreciably different than at equilibrium. The same is valid 
also for correlations of order higher than $2$, even in 
canonical simulations or if 
the correlations are defined by averaging over different 
dynamical realizations. 
}
\label{fig_correlations}
\end{figure}

On the other hand, we take here the opportunity 
of pointing out that the dynamical behavior during both
microcanonical and canonical QSS's is such that the inter-particle
correlation is negligible (see Fig. \ref{fig_correlations}).
Thus, in both cases the statistical mechanics description of
the QSS's has to be based on the assumption of independence among
elementary components  and on the consequent application of the
central limit theorem. 
This is consistent with our approach in \cite{hmf_qss}, where
we give compelling evidence of the applicability of Kinchin's 
derivation of statistical mechanics 
\cite{kinchin}, once an appropriate estimation of the density of
states $\omega(E)$ for the QSS's is given.
It is also pertinent to add that the evidences we are presenting
in Fig. \ref{fig_correlations} 
provide full justification
to the use of the Vlasov theory for the QSS's, as it is done,
e,g., in \cite{ruffo}. 

The statement \cite{tsallis} 
that the non-Gaussianity of the one-body angular
momentum PDF ``\ldots excludes the Boltzmann-Gibbs exponential form as
the energy distribution in full phase space $\Gamma$ \ldots''
is wrong. 
The Authors of \cite{tsallis} are trivially proving that the PDF of
a  {\it microstate}  $(l_i,\theta_i)_{i=1,2,\ldots,M}$ is not 
proportional to   $e^{-\beta\;H(l_i,\theta_i)}$, $H$ being the
Hamiltonian of the system with $M$ particles. 
However, this does not imply that the {\it total energy} PDF must be
different from $p(E)=\omega(E) e^{-\beta E}/Z$, where $\beta$ is an
inverse temperature.  
On the contrary, 
in force of the central limit theorem $p(E)$  still
has the usual Boltzmann-Gibbs form \cite{kinchin}, as we clearly
verify in \cite{hmf_qss}. The nontrivial point is that $\omega(E)$ for
the canonical QSS's is a {\it nonequilibrium} density of state which
can be calculated by considering a submanifold of the
$\Gamma$-space at constant magnetization \cite{hmf_qss}. 

In their final remark, the Authors of \cite{tsallis} propose a
non-exponential fitting of our results for $\ln[p(E)/\omega(E)]$ after
the introduction of an {\it ad hoc}
energy-shift $E\mapsto E-692$ for which they provide no explanation. 
A plot of the non-exponential function in
\cite{tsallis} without this energy-shift does not agree at all
with the data from the dynamical simulations. 
We also point out that by fitting
$\ln[p(E)/\omega(E)]$ one implicitly assumes the validity of 
our calculation of  
$\omega(E)$, which is based on  the
fundamental thermodynamic relation linking temperature to the
Boltzmann expression for the entropy, 
$S\equiv k_B\ln[\omega(E)]$.
As we explain in \cite{hmf_qss}, 
to propose an alternative to the exponential weight $e^{-\beta E}$
without appropriately changing the determination of $\omega(E)$ is
logically inconsistent. 

In summary, we have shown that the statements in
\cite{tsallis} either are wrong or do not apply to our 
results in \cite{hmf_qss}.

\end{document}